\documentclass[pra,twocolumn,amsmath,amssymb]{revtex4-1}
\usepackage{graphicx}

\begin{document}

To appear in a special issue of pss(b)\\

\title{Nodal Semimetals : A Survey on Optical Conductivity}

\author{A. V. Pronin}
\email{artem.pronin@pi1.physik.uni-stuttgart.de}
\author{M. Dressel}
\affiliation{1. Physikalisches Institut, Universit\"at Stuttgart,
70569 Stuttgart, Germany}

\date{March 23, 2020}

\begin{abstract}

Among different topological and related phases of condensed matter,
nodal semimetals occupy a special place -- the electronic band
topology in these materials is related to three-dimensional bulk,
rather than to surface, states. A great variety of different
realizations of electronic band crossings (the nodes) leads to a
plethora of different electronic properties, ranging from the chiral
anomaly to solid-state realizations of a black-hole horizon. The
different nodal phases have similar low-energy band structure and
quasiparticle dynamics, which both can be accessed experimentally by
a number of methods. Optical measurements with their large
penetration depth and high energy resolution are ideally suited as
such a bulk probe; especially at low energies where other
spectroscopic methods often lack the required resolution. In this
contribution, we review recent optical-conductivity studies of
different nodal semimetals, discuss possible limitations of such
measurements, and provide a comparison between the experimental
results, simple theoretical models, and band-structure-based
calculations.

\end{abstract}

\maketitle

\section{Introduction}

Succeeding graphene and topological insulators, nodal semimetals
came into focus of condensed matter physics a few years
ago~\cite{Murakami2007, Kobayashi2007, Wan2011, Burkov2011PRL,
Burkov2011PRB, Park2011, Manes2012, Wang2012, Wang2013, Liu2014,
Borisenko2014, Neupane2014, Vafek2014, Lv2015, HuangSM2015,
Shekhar2015, Weng2015, Xu2015TaAs, Xu2015NbP, Wang2016, Chiu2016,
Zhu2016, Bradlyn2016, Armitage2018}. In these there-dimensional (3D)
materials, linearly dispersing electronic bands possess point and/or
line crossings in the vicinity of the chemical potential in the bulk
Brillouin zone (BZ). These bulk band crossings may lead to
topologically trivial (as in the case of Dirac semimetals) or
nontrivial (e.g., in Weyl semimetals) electronic phases. Important
is that the low-energy electronic dispersion relations can be
approximated by a solution of Dirac equation or its
modifications~\cite{Wehling2014}. This makes the optical (interband)
response of nodal semimetals generally different from the response
of ``ordinary'' 3D metals and semiconductors and often allows
probing the low-energy band structure via optical conductivity
measurements. The studies on nodal semimetals, where the linear
frequency-dependent conductivity, $\sigma(\omega) =
\sigma_{1}(\omega) + i\sigma_{2}(\omega)$, had been measured, were
in focus of many recent experimental reports. Such measurements
reflect the bulk material properties, as the skin depth is typically
above a few tens of nanometers for any measurement frequency and of
the order of 100 nm to 1 $\mu$m for the most interesting
far-infrared portion of the spectrum~\cite{Schilling2017a,
Neubauer2018, Hutt2018}. In this paper, we summarize our findings
obtained in such measurements within the last few years and review
the most relevant optical results from literature.

In the discussion, we concentrate exclusively on the
optical-conductivity features related to the electronic band
structure. In addition, optical spectra may contain information on
such effects as strong electron-electron or electron-phonon
coupling. Possible importance of these interactions in different
Dirac materials is widely debated~\cite{Witczak2014, Dzero2016,
Liu2016b, Fujioka2017, Ye2018, Rinkel2017, Hui2019}. It looks like
the majority of experimental results on nodal semimetals
(particularly, for nonmagnetic systems) can be understood within a
single-particle picture. Still, there are reports on experimental
detection of different collective effects, also by optical means.
For example, a density-wave formation was suggested in the Dirac
semimetal Ca$_{1-x}$Na$_{x}$MnBi$_{2}$ \cite{Corasaniti2019} and a
strong coupling between optical phonons and Weyl quasiparticles was
discussed in TaAs \cite{Xu2017}. Reviewing such effects is not the
scope of this paper. Here, we just note that in all examples
discussed later the collective effects do not manifest themselves in
optical conductivity in an explicit way. For instance, the phonon
modes remain sharp (and of Lorentzian shape), even though these
modes overlap in frequency with the interband (Drude) response of
conducting carriers, see, e.g., Refs.~\cite{Neubauer2018,
Maulana2019}.

\section{Theoretical background: electronic band dispersion and optical conductivity}

\begin{figure}[]
\centering \vspace{0.3 cm}
\includegraphics [angle=-90, width=\columnwidth]{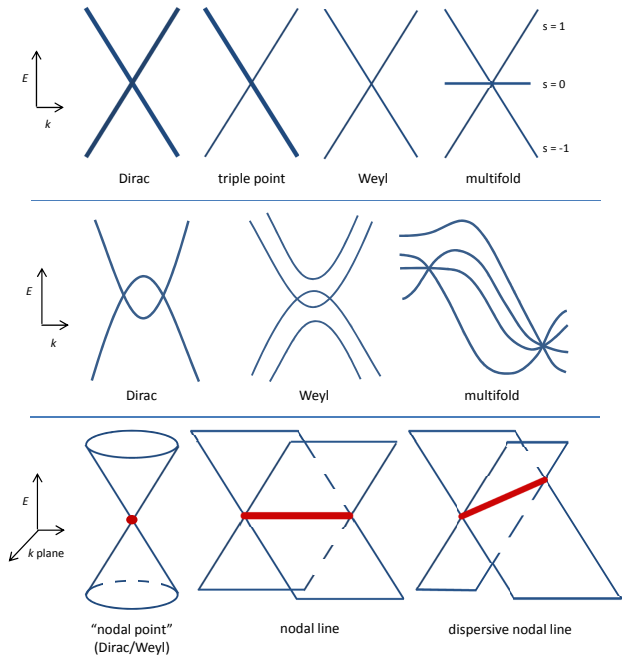}
\vspace{0.3 cm} \caption{Model band structures for different nodal
semimetals. Upper row: schematic band dispersions near the nodes in
Dirac, triple-point, Weyl, and multifold semimetals (from left to
right). The multifold-semimetal dispersion is presented for a spin-1
threefold-fermion case. Middle row: more realistic (but still
schematic) band structures, related to the compounds considered in
this review -- Dirac (Cd$_{3}$As$_{2}$), Weyl (the TaAs family), and
multifold semimetals (RhSi and other materials from the space group
No. 198). For the latter case, the electronic bands are shown
without SOC being included (cf. the accurate band-structure
calculations for RhSi in Fig.~\ref{RhSi_2}). For a triple-point
realistic band dispersion, refer to Fig.~\ref{GdPtBi}. In the two
upper rows, the band degeneracy is encoded as the line thickness --
thin lines represent nondegenerate bands, whereas thick lines
represent (doubly) degenerate bands. Bottom raw: a graphical
explanation of the differences between the Dirac (or Weyl)
semimetals (left picture) and nodal-line semimetals (middle and
right pictures). The band crossings are depicted as bold red points
and lines. Note that the actual $\mathbf{k}$-space is 3D; thus, the
shown figures are the relevant cuts of the full (four-dimensional)
$E(\mathbf{k})$ pictures.} \label{model_bands}
\end{figure}

In this section, we briefly recap theoretical predictions for the
frequency behavior of optical conductivity in the major types of
nodal semimetals (see Fig.~\ref{model_bands}). More details for
specific cases can be found in vast available literature, especially
in the works of J. P. Carbotte and coauthors~\cite{Ashby2014,
Carbotte2016, Tabert2016a, Tabert2016b, Carbotte2017,
Mukherjee2017a, Mukherjee2017b, Mukherjee2018, Carbotte2019a,
Carbotte2019b}.

In the case of electron-hole symmetric $d$-dimensional bands with
$E(\mathbf{k}) \propto \pm \lvert \mathbf{k} \rvert ^{z}$, the real
part of the interband optical conductivity is supposed to follow a
power-law frequency dependence~\cite{Hosur2012, Bacsi2013},
\begin{equation}
\sigma_{1}^{\rm{IB}}(\omega) \propto \omega ^{(d - 2)/z}.
\label{exponent}
\end{equation}
For Dirac and Weyl semimetals, $d=3$, the bands are conical ($z=1$),
and Eq.~\ref{exponent} can be more specifically rewritten as
\begin{equation}
\sigma_{1}^{\rm{IB}}(\omega) = \frac{e^2 N_{W}} {12 h} \frac{\omega}
{v_F}, \label{Weyl}
\end{equation}
where $N_{W}$ is the number of Weyl nodes (for a single Dirac node,
$N_{W}=2$), $v_F$ is the Fermi velocity, $h=2\pi\hbar$ is the Planck
constant, and all Weyl/Dirac bands are considered to be identical
(up to a spin degree of freedom) with their nodes situated at the
chemical potential $\mu$. If the node position is not at the
chemical potential ($\mu \neq 0$), transitions for the energies
below $2\mu$ are Pauli blocked, and Eq.~(\ref{Weyl}) is modified to
\begin{equation}
\sigma_{1}^{\rm{IB}}(\omega) = \frac{e^2 N_{W}} {12 h} \frac{\omega}
{v_F} \theta\left\{\hbar\omega-2\mu\right\}, \label{Heaviside}
\end{equation}
where $\theta\{x\}$ is the Heaviside step function and any carrier
scattering is ignored. In this case, an intraband contribution to
conductivity will also be present in the spectra. For finite
electron scattering, the Heaviside function can be replaced, for
example, by
\begin{equation}
\frac{1}{2}+\frac{1}{\pi}\arctan \frac{\omega-2\mu/\hbar}{\gamma}
\label{arctangent}
\end{equation}
with $\gamma$ representing an appropriate scattering rate, and
intraband conductivity can be approximated by a standard Drude
ansatz~\cite{Dressel2002}.

In Ref.~\cite{Carbotte2016}, it was shown that tilting the conical
bands (relevant, e.g., for type-II Weyl
semimetals~\cite{Soluyanov2015, Xu2015PRL}) affects the linear
behavior of optical conductivity: $\sigma_{1}^{\rm{IB}}(\omega)$
remains (quasi)linear, but experiences slope changes at certain
frequency points, whose positions are related to $\mu$ and to the
tilt angle.

For generalizations of Weyl bands with higher Chern
numbers~\cite{Manes2012, Bradlyn2016, Xu2011, Fang2012, Huang2016,
Singh2018}, the shape of $\sigma_{1}^{\rm{IB}}(\omega)$ depends on
the band dispersion relations. In the so-called multifold
semimetals, where a few linear (rotationally symmetric) bands with
generally different slopes cross at a given point of the
BZ~\cite{Manes2012, Bradlyn2016}, the optical conductivity is linear
in frequency (up to the steps, related to the Pauli-blocked
transitions)~\cite{Grushin2019}. For more complicated band
structures, such as touching bands with a linear dispersion in one
direction and parabolic dispersions in the remaining
two~\cite{Xu2011, Fang2012}, $\sigma_{1}^{\rm{IB}}(\omega)$ is
expected to be anisotropic~\cite{Ahn2017multi}, in accordance with
Eq.~\ref{exponent}. Additionally, if the nodes are situated at
different energies, as appears, e.g., in real multifold semimetals,
$\sigma_{1}^{\rm{IB}}(\omega)$ changes its frequency run at
different energy scales. The important point is that the total
interband $\sigma_{1}(\omega)$ can often be decomposed into
contributions from the nodes of each kind, simplifying
interpretation of experimental spectra.

A particularly interesting case is the nodal-line semimetals
(NLSMs)~\cite{Burkov2011PRB}, where the presence of a continues line
of nodes effectively reduces the dimensionality of the crossing
electronic bands to $d=2$. This reduced dimensionality leads to a
frequency-independent $\sigma_{1}^{\rm{IB}}(\omega)$ according to
Eq.~\ref{exponent}. Earlier, such ``flat'' optical conductivities
were predicted and experimentally observed in graphene and
graphite~\cite{Ando2002, Mak2008, Kuzmenko2008} with a universal
conductance value per one graphene sheet, $\pi e^{2}/(2h)$. In
NLSMs, no universal sheet conductance is expected; instead
$\sigma_{1}^{\rm{IB}}(\omega)$ is related to the length of the nodal
line $k_{0}$ in a BZ~\cite{Carbotte2017, Mukherjee2017a, Ahn2017}.
For a circular nodal line, one has:
\begin{equation}
\sigma_{1}^{\rm{IB}}(\omega) = \frac{e^2 k_{0}} {16 \hbar}.
\label{k_0}
\end{equation}
It is assumed here that the plane of the nodal circle is
perpendicular to the electric-field component of the probing
radiation and that there is no particle-hole asymmetry. For $\mu
\neq 0$, a Pauli edge (Eqs.~\ref{Heaviside} and \ref{arctangent})
occurs in the conductivity spectra.

\section{Linear optical response: Review of experimental results}

\subsection{Experiment versus computations}

\begin{figure}[b]
\centering
\includegraphics [width=8 cm]{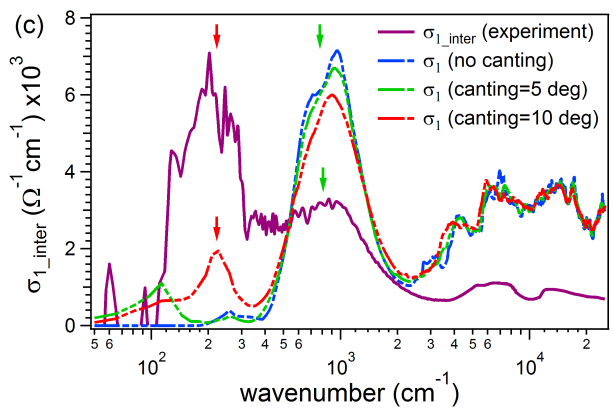}
\caption[]{Optical conductivity of YbMnBi$_{2}$. The solid magenta
line shows experimental interband conductivity (i.e., the Drude
contribution is subtracted) at 5 K. The other curves are
band-structure-based calculations for different canting angles of
Mn$^{2+}$ magnetic moments. Reproduced with permission from
Ref.~\cite{Chaudhuri2017}, copyright (2017) by the American Physical
Society.} \label{YbMnBi2}
\end{figure}

\begin{figure}[b]
\centering
\includegraphics [width=8 cm]{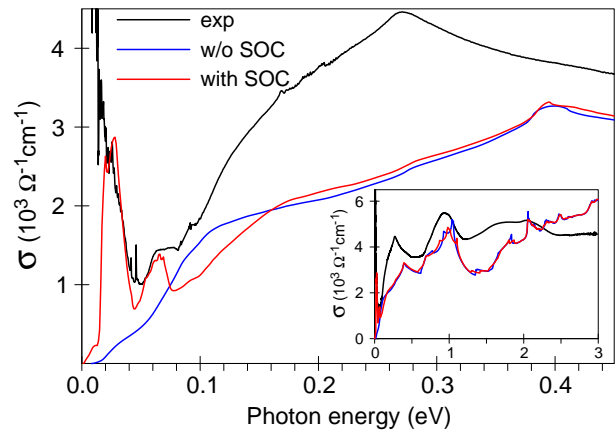}
\caption{Interband optical conductivity of NbP calculated from its
band structure with (red line) and without (blue line) SOC and the
total (i.e., inter- and intraband) experimental NbP conductivity at
10 K (black line). Intraband (Drude) contributions to the
conductivity are not included in the computations. Inset shows same
sets of data on a broader frequency scale. All spectra are for the
(001)-plane response. Reproduced with permission from
Ref.~\cite{Neubauer2018}, copyright (2018) by the American Physical
Society.} \label{NbP1}
\end{figure}

A large number of recent experiments are devoted to measuring
optical conductivity in nodal-semimetal
candidates~\cite{Schilling2017a, Neubauer2018, Hutt2018, Ueda2012,
Timusk2013, Chen2015, Sushkov2015, Xu2016, Chinotti2016,
Neubauer2016, Schilling2017b, Qiu2018, Kemmler2018, Xu2018,
Corasaniti2019, Martino2019, Qiu2019a, Qiu2019b, Maulana2019}. There
are also theoretical studies, where $\sigma(\omega)$ is computed for
particular semimetal compounds based on their band
structure~\cite{Grassano2018a, Grassano2018b, Habe2018, Habe2019,
Li2019}. Some studies combine both, experiment and
band-structure-based computations~\cite{Neubauer2018, Hutt2018,
Chaudhuri2017, Ebad2019, Shao2019}. We would like to stress here
that usually the match between measurements and such computations is
only qualitative. In Figs.~\ref{YbMnBi2} and \ref{NbP1}, we show two
typical examples of the theory-vs-experiment spectra comparison --
for YbMnBi$_{2}$ and NbP, correspondingly. In both cases, the
calculations reproduce the major features observed in the
experimental conductivity, but fail to catch the exact frequency
positions of the features and their spectral shapes. This result is
not surprising, considering the well-known difficulties of ab initio
optical-conductivity calculations, especially at low frequencies. It
seems optimal to combine experimental studies with both,
simple-model (or effective-Hamiltonian) approaches and ab initio
calculations, as attempted, e.g., in Refs.~\cite{Maulana2019,
Chaudhuri2017, Ebad2019, Shao2019}. This may allow a deeper insight
into the relation between the semimetal band structure and its
optical conductivity. Additionally, band-selective
optical-conductivity calculations are quite helpful. Such
calculations are, however, rarely performed~\cite{Neubauer2018,
Chaudhuri2017, Ebad2019}.

\subsection{Nodal-line semimetals}

\textit{\textbf{ZrSiS and its relatives.}} We start reviewing
optical-conductivity measurements of nodal semimetals with ZrSiS.
The reason for this is a relatively simple, basically model, band
structure of this layered, quasi-two-dimensional, compound. (We note
that all currently available optical measurements of ZrSiS and its
relatives were performed on in-plane surfaces and showed no
anisotropy, consistent with the tetragonal in-plane symmetry of
these compounds.) ZrSiS possesses a nodal line, situated near the
Fermi level \cite{Schoop2016, Neupane2016}. Although the shape of
the line is rather complex and, furthermore, the line is slightly
gapped due to spin-orbit coupling (SOC), the linearity of the
electronic bands forming this nodal line extends up to $\sim$0.5~eV,
and other (nonlinear) bands do not cross the Fermi level. This makes
ZrSiS one of the best systems for searching the signatures of Dirac
electrons in the optical conductivity spectra.

\begin{figure}[t]
\centering
\includegraphics [width=8 cm]{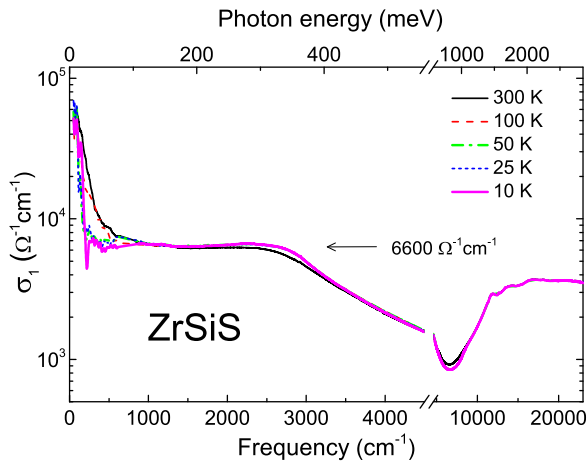}
\caption[]{In-plane optical conductivity of ZrSiS. Reprinted with
permission from Ref.~\cite{Schilling2017a}, copyright (2017) by the
American Physical Society.} \label{ZrSiS}
\end{figure}

Measurements of the optical conductivity in ZrSiS have been reported
in Refs.~\cite{Schilling2017a, Ebad2019, Ebad2019ZrSiTe, Uykur2019}.
In Fig.~\ref{ZrSiS}, we display the real part of the optical
conductivity obtained in Ref.~\cite{Schilling2017a}. The striking
feature of the spectra is the flat, frequency-independent, region
spanning from 250 to 2500 cm$^{-1}$ (30 -- 300 meV) for almost all
temperatures investigated (at $T \geq 100$ K, the flat region starts
at a bit higher frequencies because of a broadened free-electron
Drude mode). This observation is in perfect agreement with the
simple-model predictions for NLSMs discussed above
(Eqs.~\ref{exponent} and \ref{k_0}).

The sharp dip in the 10-K spectra could be interpreted as either a
Pauli edge or the spin-orbit gap, enabling hence the upper estimate
for the gap of around 30 meV. Our later magneto-optical
investigations provide a more accurate value of 26
meV~\cite{Uykur2019}, in a reasonable agrement with the calculated
value of 15 meV~\cite{Schoop2016}.

As frequency gets higher than 400 meV, $\sigma_{1}(\omega)$ first
decreases and then increases again, demonstrating a U-shape
behavior. Similar behavior was also observed in a number of related
compounds -- ZrSiSe, ZrGeS, and ZrGeTe \cite{Ebad2019} -- and
reproduced in band-structure-based calculations for all four
materials~\cite{Habe2018, Ebad2019}. Interpretation of the
high-energy upturn of the U-shaped conductivity is rather
straightforward -- it is due to transitions between almost parallel
bands near the $X$ and $R$ points of the BZ of these compounds.

According to Ref.~\cite{Ebad2019}, the low-energy part of the
U-shaped conductivity, where $\sigma_{1}(\omega)$ is roughly
proportional to $1/\omega$, can be interpreted as being due to a
``nodal plane'' (cf. Eq.~\ref{exponent} for Dirac bands and $d=1$).
In other words, there is a further electronic-band dimensionality
reduction due to the quasi-two-dimensional (band) structure of these
compounds: the dispersion is linear only in one $\mathbf{k}$-space
direction, whereas it is almost absent along the nodal line, as well
as in the out-of-plane direction. This nodal-plane picture can only
work for relatively high frequencies, as otherwise the band
structure cannot be approximated as two dimensional. Indeed, a
basically frequency-independent $\sigma_{1}^{\rm{IB}}(\omega)$ is
observed at low frequencies (below $0.4 - 0.5$ eV) in all four
materials, albeit the frequency span of the flat conductivity is
largest in ZrSiS.

These flat areas of $\sigma_{1}^{\rm{IB}}(\omega)$ cannot be
accurately reproduced by the available calculations based on the
band structure~\cite{Habe2018, Ebad2019}. Instead, the calculated
low-energy $\sigma_{1}^{\rm{IB}}(\omega)$ is found to increase with
frequency. Habe and Koshino~\cite{Habe2018} suggested that the
observed flat conductivity might be a cumulative effect of the
increasing interband conductivity and a decreasing Drude
contribution. However, the experimental data do not support this
explanation. As one can see from Fig.~\ref{ZrSiS}, the Drude term is
very narrow and does not overlap with the flat region. Also, it is
the \textit{interband} conductivity (i.e., the conductivity after
subtraction of the Drude modes), which shows the almost flat regions
at low energies in Ref.~\cite{Ebad2019}.

Overall, the flat interband conductivity of ZrSiS at low energies is
a robust experiential result (in ZrSiSe, ZrGeS, and ZrGeTe, similar
behavior is observed). The simple interpretation based on
Eqs.~\ref{exponent} and \ref{k_0} offers a good qualitative
interpretation of this result. More advanced band-structure-based
calculations of optical conductivity are required to provide a full
description for these observations.

\textit{\textbf{NbAs$_{2}$.}} This material is another example of a
NLSM. Unlike ZrSiS, the nodal lines in NbAs$_{2}$ do not form closed
loops or cages in a BZ, but span from one BZ to another. Most
importantly, the nodal lines in NbAs$_{2}$ are ``dispersive'',
meaning that the nodal-line energy position depends on the momentum.
In fact, the nodal lines in ZrSiS and its relatives also possess
such dispersion. However, it is much weaker than in NbAs$_{2}$ and
does not seem to affect the optical spectra appreciably. Similarly
to ZrSiS, the nodal lines in NbAs$_{2}$ are gapped. Shao \textit{et
al.}~\cite{Shao2019} found in NbAs$_{2}$ experimentally and also
showed analytically that the optical conductivity due to the
transitions between the linear bands, crossing along such dispersive
nodal lines, demonstrates a linear-in-frequency, rather than a
frequency-independent, behavior.

\begin{figure}[b]
\centering
\includegraphics [width=8 cm]{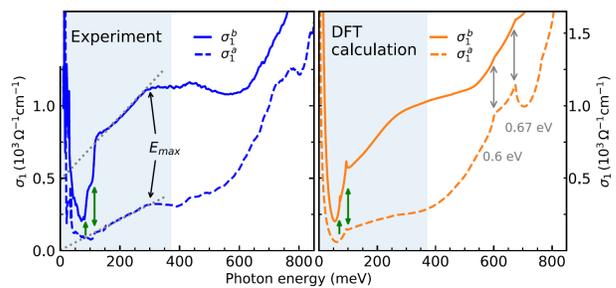}
\caption[]{Optical conductivity of NbAs$_{2}$ for two different
crystallographic directions. Left panel -- experiment, right panel
-- DFT calculations. Reproduced with permission from
Ref.~\cite{Shao2019}, copyright (2019) by the authors.}
\label{NbAs2}
\end{figure}

Indeed, the slope of the nodal line (i.e., $\partial E / \partial
k_{\|}$, where $k_{\|}$ represents the $k$ direction along the nodal
line) plays the same role as the Fermi velocity of a linear band.
Thus, for a band with a relatively large $\partial E / \partial
k_{\|}$ there will be no dimensionality reduction and the band can
be considered as a 3D Dirac band with anisotropic Fermi velocity. If
$v_{\|} = \partial E /\partial k_{\|}$ is not negligible, but still
much smaller than the Fermi velocities in the directions
perpendicular to the nodal line, $v_{\|}$ will mostly be responsible
for the slope of $\sigma_{1}^{\rm{IB}}(\omega)$. The linear increase
of $\sigma_{1}^{\rm{IB}}(\omega)$ is limited in frequency by the
energy, corresponding to the difference between the extrema of the
nodal-line energy positions. Above this frequency,
$\sigma_{1}^{\rm{IB}}(\omega)$ becomes frequency-independent (if the
bands forming the nodal line retain their linearity at these
energies).

\begin{figure}[b]
\centering
\includegraphics [width=7 cm]{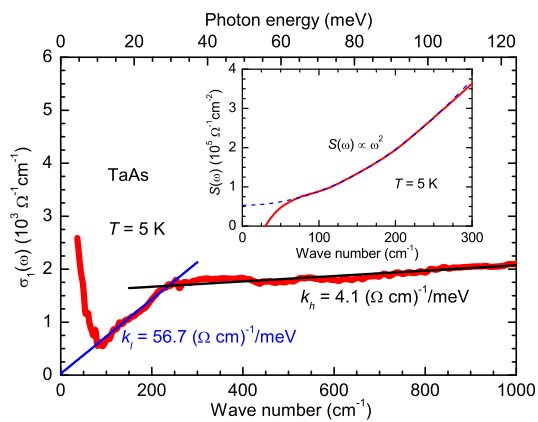}
\caption[]{Optical conductivity of TaAs. The inset shows the optical
spectral weight,
$S(\Omega)=\int_{0}^{\Omega}{\sigma_{1}(\omega)d\omega}$. The
quadratic $S(\Omega)$ corresponds to $\sigma_{1}\propto\omega$.
Reproduced with permission from Ref.~\cite{Xu2016}, copyright (2016)
by the American Physical Society.} \label{TaAs}
\end{figure}

Such behavior of the optical conductivity was recorded in
NbAs$_{2}$, see Fig.~\ref{NbAs2}. The optical conductivity is
anisotropic, because the nodal lines span almost parallel to the $a$
axis. The linear behavior of $\sigma_{1}^{\rm{IB}}(\omega)$ is
clearly seen in the measurements and can be nicely reproduced by DFT
calculations. Perhaps, it is the best match between experiment and
ab initio calculations reported for a nodal semimetal so far.

\subsection{Weyl and Dirac semimetals}

\textit{\textbf{The TaAs family.}} TaAs was one of the first
confirmed Weyl semimetals \cite{Lv2015, HuangSM2015, Yang2015} and
its optical conductivity was reported as early as in 2016
\cite{Xu2016}. This compound, as well as its family members (TaP,
NbAs, and NbP), possesses 24 Weyl nodes, i.e., twelve pairs of the
nodes with opposite chiralities \cite{Lv2015, HuangSM2015, Weng2015,
Lee2015}. The nodes are ``leftovers'' of nodal rings, which are
gapped by SOC everywhere in BZ, except of these special points. The
nodes can be divided in two groups, commonly dubbed as W1
($N_\text{W1} = 8$) and W2 ($N_\text{W2} = 16$). According to
band-structure calculations, in TaAs the W1 (W2) nodes are situated
around $20-25$~meV ($10$~meV) below the Fermi
level~\cite{Grassano2018b, Lee2015}.

\begin{figure}[t]
\centering
\includegraphics [width=7 cm]{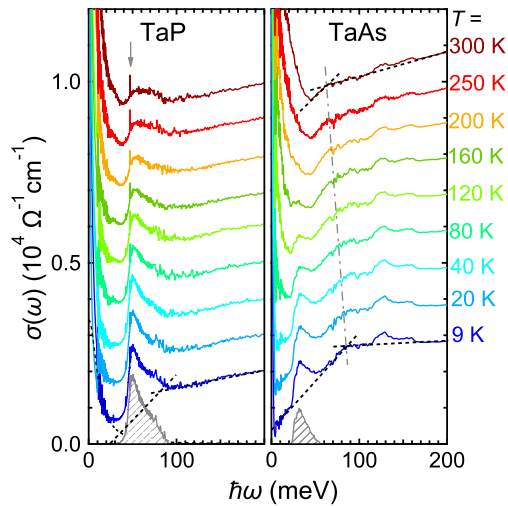}
\caption[]{Optical conductivity of TaP and TaAs. The
linear-in-frequency components of $\sigma_{1}(\omega)$ are best seen
at the lowest temperatures. The bumps at low energies might be
similar to the bumps observed in NbP (cf. Figs.~\ref{NbP1} and
\ref{NbP2}). Reproduced with permission from Ref.~\cite{Kimura2017},
copyright (2017) by the American Physical Society.} \label{Kimura}
\end{figure}

\begin{figure}[b]
\centering
\includegraphics [width=6.5 cm]{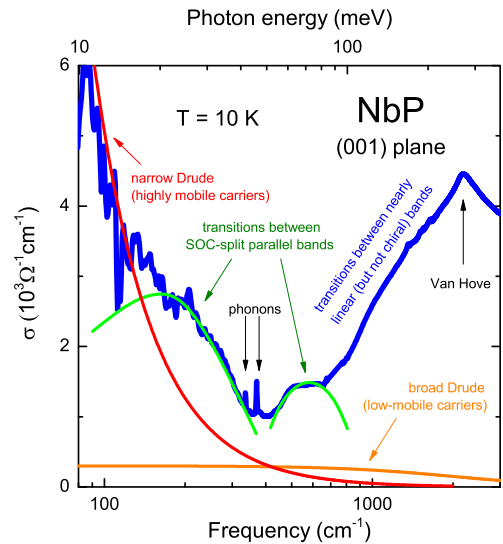}
\caption[]{Optical conductivity of NbP and assignment of the
observed features to different absorption mechanisms according to
band-selective calculations. The conductivity contributions of these
mechanisms are shown schematically as solid lines. Note logarithmic
x-scale. Reproduced with permission from Ref.~\cite{Neubauer2018},
copyright (2018) by the American Physical Society.} \label{NbP2}
\end{figure}

\begin{figure*}[t]
\centering
\includegraphics [width=15 cm]{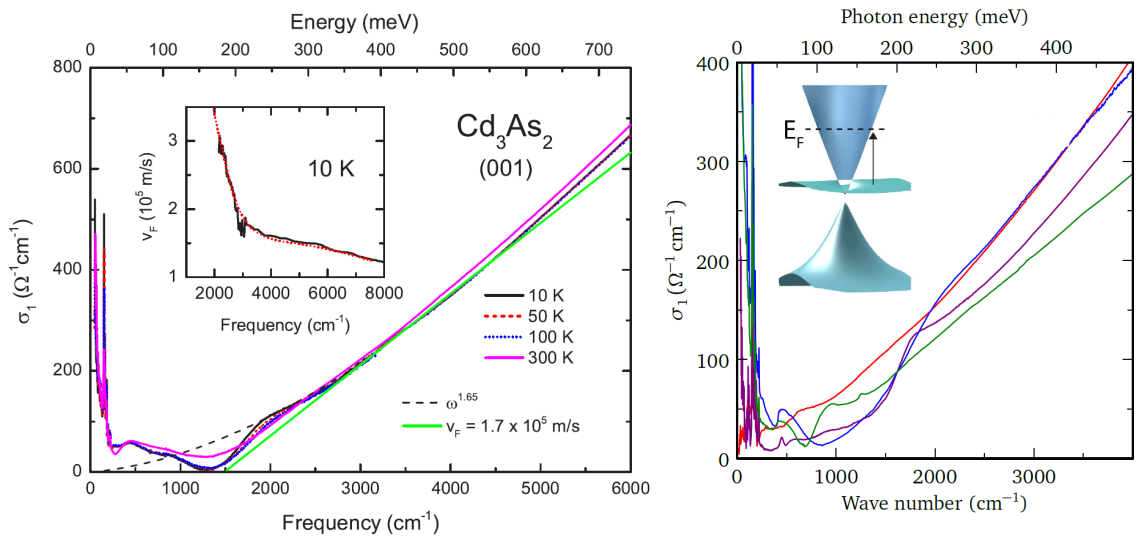}
\caption[]{Optical conductivity of Cd$_{3}$As$_{2}$ studied on a few
different samples. The graphs are adapted with permission form
Refs.~\cite{Neubauer2016} (left panel) and \cite{Crassee2018PRB}
(right panel), copyright (2016, 2018) by the American Physical
Society. In the latter study, optical response from the (112) and
(001) surfaces of different samples was collected, as shown by the
red and purple curves [the (112)-surface response] and by the blue
and green curves [the (001) response]. The inset in the left panel
demonstrates the $k$-averaged $\partial E / \partial k$, expressed
as an energy-dependent Fermi velocity, for the (112)-plane
measurements shown in this panel. Schematic Kane-model-like band
structure, proposed for Cd$_{3}$As$_{2}$ as an alternative of Dirac
cones, is depicted in the right panel.} \label{Cd3As2}
\end{figure*}

Fig.~\ref{TaAs} reproduces the real part of TaAs optical
conductivity obtained in Ref.~\cite{Xu2016}. The
zero-frequency-centered Drude mode is followed by an almost linear
$\sigma_{1}(\omega)$, which changes its slope at around 25 meV. The
low-frequency linear $\sigma_{1}(\omega)$ can be associated with the
transitions between the linearly dispersing bands near the
low-energy (W2) nodes. According to the spectra, the Pauli edge for
these transitions should be situated at or below 10 meV, implying
that the nodes are 5 meV off the Fermi level. This is not in
disagreement with the band-structure calculations mentioned above:
the accuracy of such calculations and the exact position of the
Fermi level in a given sample are both within plus/minus a few meV.

The decreased slope of the linear $\sigma_{1}(\omega)$ for
$\hbar\omega > 25$ meV may look puzzling at first glance: at higher
energies, the W1 nodes should start contributing to
$\sigma_{1}(\omega)$; hence it should increase more rapidly with
$\omega$. To qualitatively explain the decreasing slope, one should
recall that the Weyl nodes in TaAs are leftovers of gapped nodal
lines. The bands, forming the W2 nodes, flatten out in the $k$
direction along the nodal lines at quite small energies (the
nodal-line gap is in the tens-of-meV range). Thus, at these energies
the optical response starts to remind the nodal-line situation:
$\sigma_{1}(\omega)$ gets flattened.

One can notice that there is a small bump on the initial linear
$\sigma_{1}(\omega)$ at some 15 meV. This bump was reproduced in
later optical measurements of TaAs~\cite{Kimura2017}; a similar, but
much stronger, peak was detected in TaP at comparable
energies~\cite{Kimura2017, Polatkan2019}, see Fig.~\ref{Kimura}. In
Ref.~\cite{Kimura2017} these features were attributed to the
transitions between the Lifshitz points of the bands forming the
Weyl nodes (cf. Fig.~\ref{model_bands}, middle row).

Alternatively, such bumps might be related to the transitions
between almost parallel bands split by SOC. Similar features were
observed in our optical study of NbP at approximately 30 and 60 meV
(see Fig.~\ref{NbP1}) and interpreted as being due to such
transitions, based on ab initio band-resolved optical-conductivity
calculations~\cite{Neubauer2018}. We are not aware of any
band-structure-based optical-conductivity calculations for TaAs and
TaP at low enough frequencies; thus, the correct interpretation of
the observed low-energy peaks in these compounds is still to be
found.

An interpretation of different spectral features detected in the
optical conductivity of NbP is given in Fig.~\ref{NbP2}. The
band-selective optical-conductivity computations seem to be the best
way for making such assignments.

All optical-conductivity measurements of the TaAs family compounds
discussed above have been performed on (001) surfaces, which have
tetragonal crystallographic symmetry and no optical anisotropy. The
out-of-plane response (with the electric-field component of the
probing light parallel to [001] direction) was studied in
Ref.~\cite{Levy2018} for TaAs. A linear increase of the low-energy
interband conductivity was also observed for this polarization at
the energies below 25 meV.

\textit{\textbf{Cd$_{3}$As$_{2}$.}} Cadmium arsenide is one of the
first discovered Dirac semimetals~\cite{Wang2013, Liu2014,
Borisenko2014, Neupane2014, Jeon2014}. Band-structure
calculations~\cite{Wang2013} predict two Dirac cones per BZ in this
material. Its optical conductivity has been reported in a number of
publications~\cite{Neubauer2016, Crassee2018PRB, Jenkins2016,
Akrap2016, Uykur2018} and its (magneto)-optical properties were
recently thoroughly reviewed~\cite{Crassee2018PRM}.

In our brief review, we would like to point out that the (almost)
linear-in-frequency interband conductivity is observed in
Cd$_{3}$As$_{2}$ up to very high frequencies, signaling a large
energy scale of the (quasi)linear electronic bands, see
Fig.~\ref{Cd3As2}, where results collected on five different samples
are displayed. The Pauli edge is observed in the spectra at 600 to
1700 cm$^{-1}$, depending on the sample. This large variation of the
Pauli-edge position is related to the naturally present As
vacancies, whose concentration depends on sample-growth and
annealing conditions. Additionally, free carriers in
Cd$_{3}$As$_{2}$ demonstrate nonuniform spatial distribution,
forming charge puddles with characteristic scales of 100~$\mu$m, as
demonstrated by optical microscopy~\cite{Crassee2018PRB}.

A closer inspection of the conductivity spectra reveals a slight
superlinear increase of $\sigma_{1}(\omega)$. This increase was
attributed either to electron-self-energy effects or to deviations
of the crossing bands from perfect linearity~\cite{Neubauer2016}. In
the letter case, $\partial E / \partial k$ is energy dependent, as
shown in the inset of the left panel in Fig.~\ref{Cd3As2}.

Concluding the subsection on Cd$_{3}$As$_{2}$, we note that the
picture with two Dirac bands extending up to high energies (hundreds
of meV) has been challenged by magneto-optical
measurements~\cite{Akrap2016}, which are best consistent with a
Kane-like model~\cite{Kane1957, Bodnar1977, Bodnar1977a} with three
electronic bands, one of the bands being almost flat. (Dirac cones
may still appear in this model, but on a much smaller energy scale,
see the diagram in the right panel of Fig.~\ref{Cd3As2}). The
tunneling data~\cite{Jeon2014} can be (re)interpreted based on this
model. It has been argued~\cite{Crassee2018PRB} that the
$\sigma_{1}(\omega)$ spectra are also consistent with the model.
Still, band-structure calculations and ARPES results favor a Dirac,
rather than a Kane-like, picture for Cd$_{3}$As$_{2}$. A full
consensus about the electronic band structure of this material is
still to be established.

\textit{\textbf{Semimetals with strongly tilted Dirac or Weyl
cones.}} As mentioned above, tilting the Dirac or Weyl cones should
lead to the modifications of the interband optical conductivity:
$\sigma_{1}^{\rm{IB}}(\omega)$ is still linear, but demonstrates
changes in its slope at certain frequency points. A body of
experimental work was conducted on materials with (supposedly)
strongly tilted 3D Dirac or Weyl cones~\cite{Chaudhuri2017,
Chinotti2016, Beyer2016, Frenzel2017, Kimura2019}. Linear portions
of experimental $\sigma_{1}(\omega)$ were indeed reported, e.g., for
YbMnBi$_{2}$ -- a type-II Weyl semimetal candidate. However,
intraband contribution often masks the interband optical transitions
in such materials. This is particularly relevant for type-II
semimetals, where free carriers exist even if the chemical potential
is situated at the nodal point~\cite{Zhu2016} and, hence, Drude-like
contributions are supposed to dominate the optical-conductivity
spectra. This was indeed observed, for example, in WTe$_{2}$ and
MoTe$_{2}$~\cite{Frenzel2017, Kimura2019}.

\begin{figure}
\centering
\includegraphics[width=8 cm]{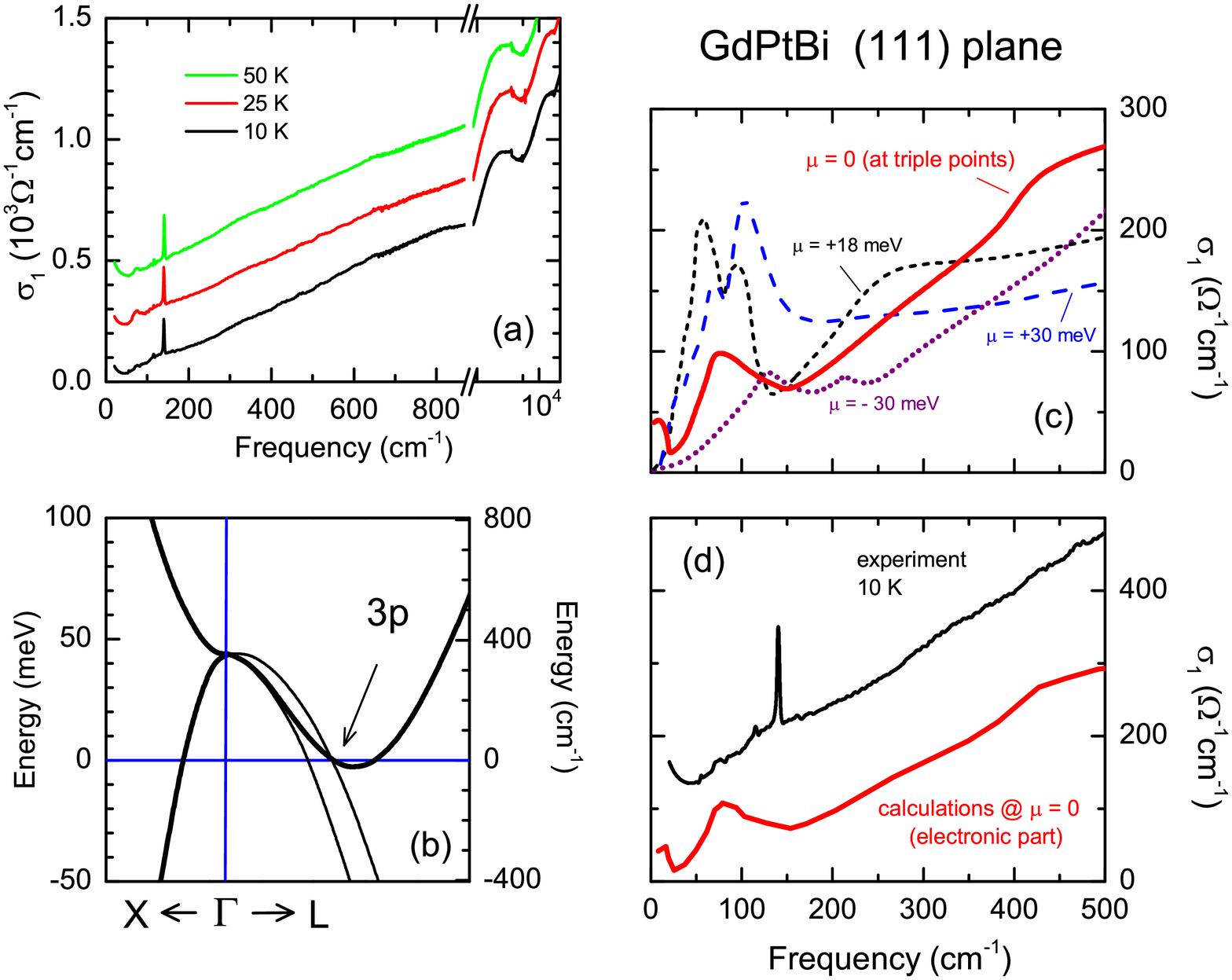}
\caption{Panel (a): Experimental optical conductivity of GdPtBi at
three different temperatures (the curves for 25 and 50 K are shifted
upwards for clarity). Panel (b): Band structure of GdPtBi near the
triple point (3p). The chemical potential is set to the triple-point
position. Doubly degenerate bands are shown as thick lines and
nondegenerate bands as thin lines. Panel (c): Calculated interband
conductivity of GdPtBi for a few different positions of $\mu$ as
indicated. Panel (d): Comparison of the measured (upper curve) and
calculated for $\mu=0$ (bottom curve) optical conductivity of
GdPtBi. The sharp phonon mode at around 150 cm$^{-1}$ is not
included in the calculated conductivity. The experimental curve is
shifted upwards by 100 $\rm{\Omega}^{-1}$cm$^{-1}$ for clarity.
Adapted with permission from Ref.~\cite{Hutt2018}, copyright (2018)
by the American Physical Society.} \label{GdPtBi}
\end{figure}

\subsection{Triple-point semimetals}

The existence of triple points, where one nondegenerate and one
doubly degenerate band cross, implies the presence of bulk nodal
lines in the band structure of triple-point semimetals
(TPSMs)~\cite{Zhu2016}. Because of this band-structure complexity,
no simple models for the optical conductivity are available for
these materials. Thus, a comparison between experiment and
band-structure-based computations is the way to interpret
$\sigma_{1}^{\rm{IB}}(\omega)$ in TPSMs. As an example, we review
below the results of our optical-conductivity measurements in
GdPtBi~\cite{Hutt2018}, a TPSM and a member of the half-Heusler
family, which is recognized for a broad variety of exotic and
potentially functional properties~\cite{Chadov2010, Lin2010}.

In the paramagnetic state (GdPtBi enters the antiferromagnetic state
at 9 K \cite{Canfield1991} -- this phase was not examined by
optics), at low temperatures (e.g., at 10 to 50 K), we found
$\sigma_{1}(\omega)$ in GdPtBi to be linear in a broad frequency
range: the linearity spans down to 100 cm$^{-1}$, see
Fig.~\ref{GdPtBi}~(a,d), indicating a low free-carrier density.
Unlike the situation in a simple conical band, this linearity is not
due to the transitions within such a band. Our calculations showed
instead that the linear $\sigma_{1}^{\rm{IB}}(\omega)$ is a
cumulative effect of transitions between a few bands with
predominantly, but not exclusively, linear dispersion relations
(Fig.~\ref{GdPtBi}~(b)). We also found that varying the position of
the chemical potential within only $\pm 30$ meV drastically changes
the overall shape of $\sigma_{1}^{\rm{IB}}(\omega)$, as demonstrated
in Fig.~\ref{GdPtBi}~(c). Thus, the simple conical dispersion, where
$\mu$ only affects the frequency position of the Pauli edge, see
Eqs.~\ref{Heaviside} and \ref{arctangent}, is obviously not relevant
for GdPtBi.

\subsection{Multifold semimetals}

Multifold semimetals -- the materials, which possess the
characteristic electronic band crossings with degeneracies higher
than two~\cite{Manes2012, Bradlyn2016} -- attract currently a lot of
attention. This electronic phase may occur in noncentrosymmetric
compounds with no mirror planes. A number of multifold semimetals
were recently predicted and experimentally confirmed, leading to a
realization of ``topological chiral crystals'' \cite{Chang2018,
Chang2017, Tang2017, Sanchez2019, Rao2019, Schroter2019,
Takane2019}. Among other chirality-related properties, these
materials are believed to demonstrate a peculiar nonlinear optical
phenomenon -- the quantized circular photogalvanic effect
(QCPGE)~\cite{deJuan2017}. In this effect, the circularly polarized
photons excite the chiral band carriers in such a way that the
resultant photocurrent is quantized in units of material-independent
fundamental constants. Recently, the observation of QCPGE was
reported in RhSi~\cite{Rees2019}, an established multifold
semimetal~\cite{Chang2017, Tang2017, Sanchez2019}. The knowledge of
frequency-dependent linear conductivity in multifold semimetals is
also essential, in particular, for a proper interpretation of QCPGE
experiments. Here, we review recent reports on experimental
determination of $\sigma_{1}(\omega)$ in RhSi~\cite{Maulana2019,
Rees2019} and compare the obtained results with the available
theoretical calculations~\cite{Grushin2019, Li2019}. As noticed
above, the optical conductivity of multifold semimetals is supposed
to demonstrate a linear-in-frequency $\sigma_{1}^{\rm{IB}}(\omega)$,
similarly to 3D Dirac and Weyl semimetals.

\begin{figure}[b]
\centering
\includegraphics [width=5 cm]{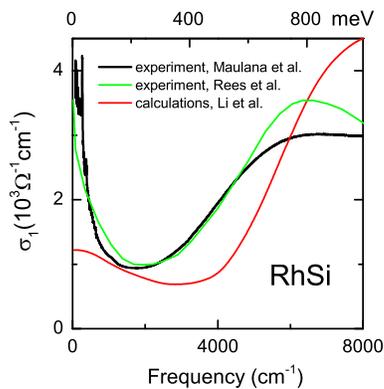}
\caption[]{Optical conductivity of the multifold semimetal RhSi. The
experimental data are from Refs.~\cite{Maulana2019} (black line) and
\cite{Rees2019} (green line). The theoretical curve is adapted from
Ref.~\cite{Li2019}.} \label{RhSi_1}
\end{figure}

Fig.~\ref{RhSi_1} shows experimental~\cite{Maulana2019, Rees2019}
and calculated~\cite{Li2019} optical conductivity of RhSi. The
experimental curves follow each other quite well. The deviations
between the curves can be explained by different free-carrier
contributions. Despite some discrepancy between the calculations and
both experimental curves, the match can be considered as
satisfactory (cf. Figs.~\ref{YbMnBi2} and ~\ref{NbP1}). Both
low-energy features of the interband experimental conductivity --
the initial (i.e., for the frequencies just above the Drude
roll-off) linear increase and the further flattening -- are
reproduced by theory.

\begin{figure}[t]
\centering
\includegraphics [width=7 cm]{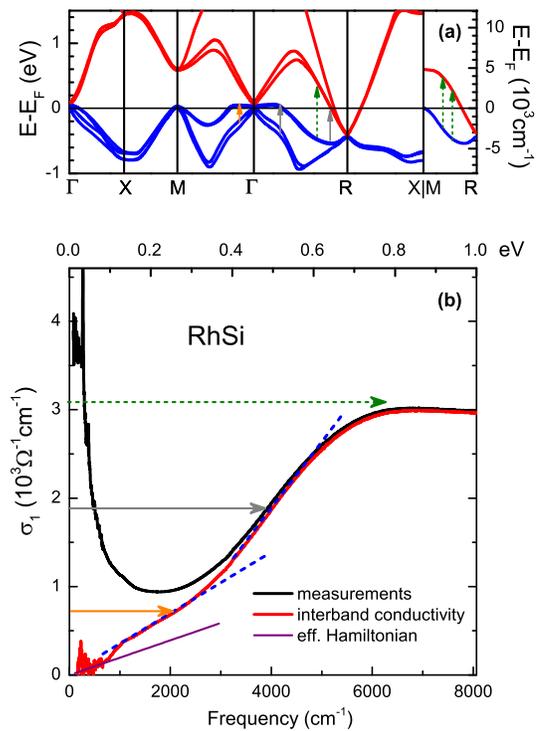}
\caption[]{Low-energy electronic structure of RhSi (a) and its
optical conductivity (b). The optical transitions responsible for
the characteristic features in the interband portion of optical
conductivity are depicted as vertical arrows in (a). The
corresponding frequency scales are indicated by the horizontal
arrows of the same color in (b). Adapted from
Ref.~\cite{Maulana2019}.} \label{RhSi_2}
\end{figure}

To establish a better connection between the features observed in
the most interesting, low-energy, part of the experimental
conductivity and the interband optical transitions, in
Fig.~\ref{RhSi_2} we show $\sigma_{1}(\omega)$ from
Ref.~\cite{Maulana2019} together with the low-energy band structure
of RhSi. Additionally, the interband contribution to the optical
conductivity, $\sigma_{1}^{\rm{IB}}(\omega)$, from this reference,
is presented. At the lowest frequencies (below approximately 2500
cm$^{-1}$), the interband conductivity is entirely caused by
transitions in the vicinity of the $\Gamma$ point. No other
interband optical transitions are possible (either the direct gap
between the bands is too large, or the transitions are Pauli
blocked). The bands near the $\Gamma$ point are all roughly linear
(two of them are basically flat); thus, a linear-in-frequency
interband conductivity is expected~\cite{Grushin2019}. Indeed,
$\sigma_{1}^{\rm{IB}}(\omega)$ is proportional to frequency in this
range (cf. the orange arrows in both panels of Fig.~\ref{RhSi_2}).
At somewhat higher frequencies ($\sim$~$3000-4000$~cm$^{-1}$), the
flat bands start to disperse downward; thus, the linearity of
$\sigma_{1}^{\rm{IB}}(\omega)$ is not expected anymore. However, the
interband contributions in the vicinity of the $R$ points become
allowed at roughly the same energy (see the grey arrows in panel
(a)). These transitions provide a dominating contribution to
conductivity, and the linear-in-frequency increase of
$\sigma_{1}^{\rm{IB}}(\omega)$ is restored with a larger slope (the
grey arrow in panel (b)). At frequencies above $6000$ cm$^{-1}$, the
optical conductivity flattens out, forming a broad flat maximum. It
can be attributed to the transitions between the almost (but not
exactly) parallel bands along the $M$--$R$ line, which are shown as
dotted green arrows. The maximum is not sharp because other
transitions with comparable energies also contribute at these
frequencies; see, e.g., the dotted green arrow between the $\Gamma$
and R points. In Fig.~\ref{RhSi_2}, we also show the results of
effective-Hamiltonian calculations~\cite{Grushin2019} for the
contributions near the $\Gamma$ point. An extrapolation of these
calculations (originally performed for frequencies below $320$
cm$^{-1}$) to higher frequencies is shown as a solid purple line.
The experimental $\sigma_{1}^{\rm{IB}}(\omega)$ is generally steeper
than the results of these calculations. The mismatch can be related
to deviations of the bands from linearity even at low
energies~\cite{Li2019, Chang2017, Tang2017}. This can be clarified
in more advanced band-structure-based optical-conductivity
calculations. In any case, the predicted linear run of
$\sigma_{1}^{\rm{IB}}(\omega)$ is experimentally confirmed for a
multifold semimetal.

\subsection{Linear-in-frequency conductivity in other materials}

The fact that 3D Dirac and Weyl semimetals were predicted to
demonstrate a quite unusual (linear-in-frequency) optical
conductivity, stimulated experimental efforts in finding such
$\sigma_{1}(\omega)$ in different materials and in making proposals
based on these observations. For example, Timusk \textit{et al.}
\cite{Timusk2013} reported a linear $\sigma_{1}(\omega)$ in a number
of quasicrystals and suggested the presence of 3D Dirac fermions in
these materials. To our knowledge, this proposition remains to be
confirmed by other experimental methods, as well as by theory. The
original forecasts of a Weyl state in pyrochlore
iridates~\cite{Wan2011} stimulated an optical study of
Eu$_{2}$Ir$_{2}$O$_{7}$, where linear $\sigma_{1}(\omega)$ was
observed in a limited range at low energies~\cite{Sushkov2015}. Up
to now, no firm confirmations of a Weyl state in this or other
pyrochlore iridates are reported; they are currently believed to be
trivial antiferromagnetic insulators with important role of electron
correlations~\cite{Nakayama2016, Wang2017a, Wang2017b}.
Interestingly to note that linear $\sigma_{1}(\omega)$ in a broad
frequency range was observed in BaCoS$_{2}$, another material with
strong electron correlations~\cite{Santos2018}. This linearity was
not attributed to 3D conical bands (which actually do not exist in
this compound), but to an effect of electron correlations. These
examples, as well as the case of GdPtBi discussed above, demonstrate
once again that interpretations of linear $\sigma_{1}(\omega)$
should always be made with care and theory output is essential for
such interpretations.

\section{Conclusions}

When graphene shifted in the focus of condensed matter physics, its
constant in frequency conductivity appeared as a peculiarity at
first glance. Soon it became clear that this was the tip of an
iceberg: materials with interesting band structure and topology are
predicted and discovered at a rapid pace since. The Dirac cones in
graphene are just a particular case of a class of systems that is
not restricted to two dimensions. Different nodal semimetals,
possessing conical bands in their bulk, are currently at the center
of studies, and conductivity of these materials is directly related
to the electronic band dispersion and dimensionality. Albeit ARPES
is surely the most proper method to study the band structure and
Fermi surface, its severe restriction to the sample surface often
causes problems that can be overcome by optical methods, as a
genuine bulk sensitive technique. In this brief survey, we
considered a representative selection of recent experimental results
on optical studies of different nodal semimetals. The presented
examples demonstrate the abilities, as well as limitations, of
linear optics to reveal the bulk band structure at low energies.

\begin{acknowledgements}
We thank the Deutsche Forschungsgemeinschaft (DFG) for financial
support.
\end{acknowledgements}

\bibliographystyle{apsrev4-1}
\bibliography{refs}

\end{document}